\documentstyle[epsbox]{mn}
\title{Emission line profiles from self-gravitating toroids around black 
holes}
\author[Fumihiko Usui, Shogo Nishida and Yoshiharu Eriguchi]
       {Fumihiko Usui$^1$, Shogo Nishida$^2$ and Yoshiharu Eriguchi$^2$ \\
        $^1$Department of Earth Science and Astronomy, College 
        of Arts and Sciences, University of Tokyo,\\ 
        Meguro, Tokyo 153, Japan \\
        $^2$Department of Earth Science and Astronomy, Graduate School 
        of Arts and Sciences, University of Tokyo,\\ 
        Meguro, Tokyo 153, Japan}
\date{Updated \today}

\begin{document}
\maketitle

\begin{abstract}

We have computed line profiles from self-gravitating toroids around black
holes. The specific angular momentum of the toroids is assumed to be constant 
in space. The images of the toroids show peculiar feature in the rear side of 
the black holes. Concerning the line profiles, the red wing extends to the 
very small frequency region because the location of the inner edge is rather 
near the event horizon of the black hole and consequently the velocity of the 
inner edge of toroids can be faster than that of the Kepler disks.

\end{abstract}

\begin{keywords}
Black holes -- Accretion disks -- Self-gravitating toroids -- Emission Line 
profiles
\end{keywords}

\section{Introduction}

Recently, strong evidence for existence of supermassive black holes in the 
central regions of galaxies has been found from observations of 
electromagnetic waves with various wavelengths ranging from the radio waves to 
X-rays (see e.g. Miyoshi et al. 1995 for the radio observation; Ford et al. 
1994, Harms et al. 1994 for the optical observation; Tanaka et al. 1995 for 
the X-ray observation).  Radio and optical observations have shown that there 
exist very rapidly rotating gaseous disks in the central regions of galaxies. 
Since the high velocity of a disk implies the existence of a large amount of 
mass within a region of a very small size, it is widely considered that there 
are supermassive black holes with masses of $10^7 \sim 10^9 M_{\sun}$ at the 
centers of galaxies. However, such observations do not reveal the nature of 
black holes because the size of the observed region is still too large to get 
detailed information about the black holes.

Contrary to these optical and radio analyses, recent X-ray observations have 
brought us important information about gravitational fields very near the 
massive objects or black holes. By using the ASCA satellite, broad iron 
emission lines have been detected in active galaxies (see e.g. Fabian et al. 
1994; Mushotzky et al. 1995). In particular, Tanaka et al.~(1995) observed 
the Seyfert 1 galaxy, MCG--6--30--15, and discovered a very broad and skewed 
iron emission line. The broadness and skewness of the line profile can be 
explained only by assuming that the inner edge of the accretion disk is 
located very near the event horizon of the central black hole (Tanaka et al. 
1995; Fabian et al. 1995). It implies that X-ray observations can be used
to understand the very vicinity of the black holes. It is important to know 
the gravitational field near the event horizon because it is the key feature to distinguish a Schwarzschild black hole from a Kerr black hole. 

Consequently we have reached a stage to investigate very strong gravitational 
fields of black holes and to be able to test the validity of general 
relativity.  However, since the emission line 
profiles depend not only on the gravitational fields but also on the
structures of the accretion disks, it is not easy to determine the 
type of the black hole, i.e. whether the central black hole is of a 
Schwarzschild type or of a Kerr type. In fact, concerning the nature of the 
black hole at the center of MCG--6--30--15, various discussions have not
settled down to a unique interpretation yet (see e.g. Tanaka et al. 1995; 
Iwasawa et al. 1996; Dabrowski et al. 1997; Bromley et al. 1997; Reynolds \& 
Begelman 1997; Bromley et al. 1998). Therefore, in order to get a consistent 
picture of black hole -- accretion disk systems, we have to pursue much more 
investigations theoretically as well as observationally. 

Concerning the theory about the spectra of accretion disks, 
Cunningham~(1975) was the first to formulate the problem for Kerr black holes 
and obtained theoretical spectra from accretion disks around black holes (see 
also Cunningham 1976).  Gerbal \& Pelat~(1981) investigated lines from a ring 
around a black hole and found double-peaked asymmetric profiles. From the end 
of 80's, observations of lines in the X-ray spectrum stimulated many authors 
to study line profiles of accretion disks as well as disk structures 
(e.g. Nandra et al. 1989; Fabian et al. 1989; Kunieda et al. 1990; Kojima 1991;
Laor 1991; Chen \& Halpern 1989, 1990).  However, in almost all theoretical 
studies mentioned above, investigations of the emission line profiles have 
been done by assuming that disks are geometrically thin and that only direct 
photons are observed.

Some authors have studied the effect of multiple images (Luminet 1978; Bao, 
Hadrava \& {\O}stgaard 1994) and that of self-eclipse due to toroidal 
configurations (Bao \& Stuchlik 1992; Kojima \& Fukue 1992).  For toroidal 
configurations the rotation law of the toroid is not always that of the Kepler 
rotation because of the presence of the pressure within the toroid. In fact, 
Kojima \& Fukue~(1992) employed a variety of rotation laws. However, their 
analysis was done in the framework of Newtonian gravity. Therefore, 
quantitative treatment of non-Keplerian toroids in general relativity has not 
been carried out yet. 

Furthermore, in some situations, self-gravity of disks or toroids
plays an important role for the structures of disks or toroids.
In particular, massive neutron toroids around neutron stars or black
holes have been proposed as possible sources of $\gamma$-ray bursts 
(e.g. Paczy\'nski 1991; Narayan, Paczy\'nski \& Piran 1992; 
Jaroszy\'nski 1993; Witt et al. 1994).  Although there might be little
chance to observe line profiles from such systems even if exist,  it would
be interesting to study the effect of self-gravity of toroids on the
line profiles.  Such self-gravitating disks were investigated by Karas, 
Lanza \& Vokrouhlicky~(1995). However, they have studied very light thin
disks whose mass is less than several percent of the mass of the black hole.
From the theoretical point of view, it would be interesting to investigate more
massive disks or toroids as well as the effect of geometrical thickness.
By treating massive toroids, the central objects are no more of 
Schwarzschild type nor Kerr type black holes because of the gravitational 
effect of the self-gravitating massive toroids on the black holes (Nishida \& 
Eriguchi 1994). Moreover, gravitational effect of massive toroids may 
bring some differences to line profiles.  Therefore, in this paper, 
we will consider self-gravitating toroid -- black hole systems and study their 
effect on emission line profiles.

\section{Method of calculation}

\subsection{Toroid -- black hole systems}

\begin{figure}
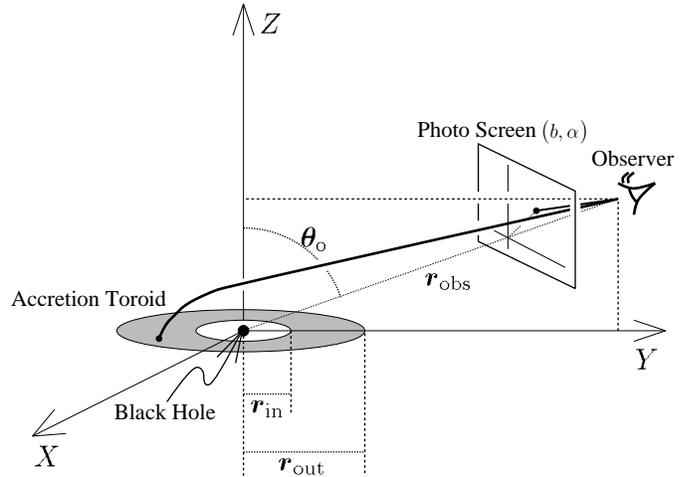

\epsfile{file=model.eps,width=0.5\textwidth}
\caption{
Schematic view of the toroid -- black hole system. The observer is located 
at $(r_{\rm obs},\theta_0)$. The coordinates of the intersection of the 
tangential line to the photon trajectory with the screen are expressed as
$(b\cos\alpha, b\sin\alpha)$. The distances to the inner and outer
edges of the toroid from the black hole are $r_{\rm in}$ and $r_{\rm out}$,
respectively.
}
\label{coordinate}
\end{figure}

We will solve the spacetime around a self-gravitating 
and geometrically thick toroid -- black hole system and study photon 
trajectories emitted from the toroid. The setting of the system is
shown in Fig.~\ref{coordinate}. The black hole is located at the origin of the 
coordinate system and the equatorial plane of the accretion toroid coincides
with the $X$-$Y$ plane. Distances from the black hole to the inner and outer 
edges of the toroid are $r_{\rm in}$ and $r_{\rm out}$, respectively. The 
position of an observer is assumed to be at $r=r_{\rm obs}$ and $\theta = 
\theta_0$ in the $Y$-$Z$ plane.  Here $(r, \theta, \varphi)$ are the 
spherical coordinates.

In order to obtain data for the images of the toroid and the line profiles,
we consider a screen perpendicular to the line connecting the black hole and 
the observer. A set of intersections of tangential lines to the photon 
trajectories with this screen forms an image of the toroid. On this screen, 
we set the polar coordinates $(b, \alpha)$ whose origin is at the intersection 
of the line connecting the black hole and the observer with the screen. 

\subsection{Metric and photon trajectories}

Since we will treat {\it stationary} and {\it axisymmetric} toroid -- black 
hole systems, the line element can be written in the pseudo--isotropic 
coordinates as follows:
%
\begin{eqnarray}
\label{line element}
ds^2 = -e^{2\psi} dt^2 + e^{2\xi}({1 \over 1-2M/r+J^2/(M r)^2}dr^2+
r^2d\theta^2)\nonumber\\
+e^{2\zeta}r^2\sin^2\theta(d\phi-\omega dt)^2 ,
\end{eqnarray}
where $t$ and $\phi$ are the coordinates associated with time and axial 
Killing vectors, respectively and $M$ and $J$ are the gravitational mass
and the total angular momentum of the black hole, respectively.
Metric components $\psi$, $\xi$, $\zeta$ and $\omega$ are functions of 
$r$ and $\theta$. This metric reduces to that of the Schwarzschild
coordinates in the spherical limit. Hereafter we use the units of 
$c=G=1$.

We employ the numerical code of Nishida \& Eriguchi~(1994) to solve
the structures of toroids and the metric functions.  The matter of the toroid
is assumed to be polytrope as follows:
\begin{equation}
p = K \rho^{1+1/N}\ , \qquad \varepsilon = \rho +Np \ ,
\end{equation} 
where $p$, $\rho$, $\varepsilon$, $N$ and $K$ are the pressure, 
the rest mass density, the energy density,
the polytropic index and a constant, respectively. We choose $N=3$ polytropes
in this paper. Concerning the rotation law of the toroid, the specific angular 
momentum $l$ is assumed to be constant, i.e. $l = $ constant. 

Trajectories of photons emitted from the surface of the toroid are followed
up to the observer. This can be done by integrating the geodesic equation
for photons:
\begin{equation}
\label{eq of tra.}
\frac{du^{\alpha}}{d\lambda}+\Gamma^{\alpha}_{\mu\nu}u^{\mu}u^{\nu}=0\ ,
\end{equation}
where $\lambda$ is an Affine parameter, $u^{\alpha}$ is the four--velocity of 
the photon which is defined by
\begin{equation}
u^{\alpha}=\frac{dx^{\alpha}}{d\lambda}\ ,
\end{equation}
and $\Gamma^{\alpha}_{\mu\nu}$ is the Christoffel symbol.

Although photons are emitted from the surface of the toroid, we integrate the 
geodesic equation starting from the observer position back to the toroid. If 
the geodesic hits the surface of the toroid, we can get one trajectory of
a photon connecting the surface of the toroid and the observer. Thus the 
geodesic equation is integrated by choosing initial values for the 
four--velocity at the observer position. Since the metric functions are 
solved on discrete mesh points, we need to interpolate the metric functions 
at points where the geodesic passes. Once a geodesic curve from the toroid to 
the observer is obtained, the position of the image on the screen can be
determined. By integrating many photon trajectories, we have
an image of the toroid on the screen and investigate the spectrum
from the toroid. In practice, we have computed $(100 \times 100)$ photon 
trajectories by choosing initial velocities so that tangential lines hit mesh 
points on the screen which are equidistantly distributed.

\subsection{Observed total intensity}

In order to calculate the observed total intensity from the toroid, we need to 
specify the distribution of emissivity of the toroid. For geometrically 
thin and nonself-gravitating disks, there are several models for emissivity
(e.g. Novikov \& Thorne 1973; Page \& Thorne 1974), but they are not decisive 
ones because of the lack of observational data. The situation is worse for 
self-gravitating toroids. Therefore, at the present stage, we adopt a 
simplified model in which the emissivity is assumed to be a power of the 
distance from the rotation axis as follows:
\begin{equation}
\label{emissivity}
e = k R\,^s\ \ ,
\end{equation}
where $e$, $k$ and $s$ are the emissivity and two constants, respectively,
and $R$ is the distance from the rotation axis defined as 
\begin{equation}
R = r \sin \theta \ .
\end{equation}

Since the toroid is rotating around the black hole, the energy of photons 
emitted from the toroid is shifted due to the Doppler effect and the
gravitational redshift by the following factor $g$:
\begin{equation}
\label{gdef}
g \equiv {1 \over 1+z} = \frac{E_{\rm obs}}{E_{\rm em}} \ ,
\end{equation}
where $z$, $E_{\rm em}$ and $E_{\rm obs}$ are the redshift, the energy of the
emitted photon and the energy of the observed photon, respectively. Since the 
matter in the toroid is assumed to move only in the azimuthal direction,
the energy of the emitted photon is expressed as 
\begin{equation}
E_{em} = p_tu^t+p_{\phi}u^{\phi} = p_tu^t
\left(1 + \Omega_t \frac{p_{\phi}}{p_t}\right)\ ,
\end{equation}
where $u^t$, $u^{\phi}$, $p_t$ and $p_{\phi}$ are the $t$-- and 
$\phi$--components of the four--velocity of the toroid and the covariant 
$t$-- and $\phi$--components of the four--momentum of the photon,
respectively. The quantity $\Omega_t$ is the angular velocity 
of the toroid which is defined as 
\begin{equation}
\Omega_t = {u^{\phi} \over u^t} \ .
\end{equation}
This angular velocity can be expressed as follows for the constant
specific angular momentum toroid:
\begin{equation}
\Omega_t = \omega + {l \over (e^{\zeta-\psi} r \sin \theta)^2 
(1 - l \omega)} \ .
\end{equation}
The component $u^t$ can be calculated from the normalization condition 
$u_{\mu}u^{\mu}=-1$ as follows:
\begin{equation}
u^t=\frac{1}{\sqrt{e^{2\psi}-e^{2\zeta}r^2\sin^2\theta\;
(\Omega_t-\omega)^2}}\ .
\end{equation}

Since photons travel in the axisymmetric and stationary spacetime,
quantities $p_t$ and $p_{\phi}$ are conserved along the photon trajectories. 
Thus, the energy of the observed photon is given as $E_{\rm obs}=p_t$ and
the ratio $1 + z =E_{\rm em}/E_{\rm obs}$ can be expressed as follows:
\begin{equation}
\label{1+z}
1+z = u^t\left(1 + \Omega_t \frac{p_{\phi}}{p_t}\right)\ .
\end{equation}
In this equation, the quantity $p_{\phi}/p_t$ is the impact parameter of 
the photon around the $Z$-axis and is expressed as
\begin{equation}
p_{\phi}/p_t=b\sin\theta_0\;\sin\alpha\ .
\end{equation}

By using this redshift factor, the observed total intensity 
$I^{(0)}_{\rm obs}$ can be calculated from the intrinsic total intensity
of the source $I^{(0)}_{\rm em}$ as follows:
\begin{equation}
I^{(0)}_{\rm obs}=\left(\frac{1}{1+z}\right)^4\;I^{(0)}_{\rm em} \ .
\end{equation}
The 4th power in this expression may be understood as (1) the Doppler effect 
by the rotation of the toroid (one power), (2) the redshift by the 
gravitational filed (one power), and (3) the focusing effect by the 
gravitational lens effect (two power) (see e.g. Luminet 1978).

\subsection{Line profiles}

The observed specific flux $dF_{\rm obs}(E_{\rm obs})$ within the solid angle 
$d\Omega$ can be calculated from
\begin{equation}
dF_{\rm obs}(E_{\rm obs}) = I_{\rm obs}(E_{\rm obs}) d \Omega 
= g^3 I_{\rm em}(E_{\rm em}) \; d\Omega \ ,
\end{equation}
where $I_{\rm obs}$ and $I_{\rm em}$ are the observed specific intensity and 
the specific intensity at the source, respectively. In this equation we have 
made use of the invariance of the quantity $I/E^3$ along the geodesics. Since 
we consider a narrow line, the specific intensity $I_{\rm em}$ in the 
rest frame of the source is assumed to be a delta function as follows:
\begin{equation}
I_{\rm em}(E_{\rm em}) = e(R) \delta(E_{\rm em}-E_0) \ ,
\end{equation}
where $e$ is the emissivity given in equation (\ref{emissivity}) and 
$E_0$ is the rest energy of the emitted photon. In this paper, the emission is 
assumed to be isotropic.

\section{Numerical results}

\subsection{Models for toroid -- black hole systems}

We have computed three equilibrium configurations of toroid -- black hole
systems and their metric components by using the general relativistic code
(Nishida \& Eriguchi 1994): models with the mass ratio $q \equiv m/M = 0.012$
(Model A hereafter), $0.080$ (Model B) and $0.49$ (Model C) where $m$ is 
the mass of the toroid. As mentioned in Introduction, it should be noted that 
the central object is neither a Schwarzschild nor a Kerr black hole because 
of {\it deformation} due to gravity of the toroid. It implies that the 
spacetime is different from that of Schwarzschild or Kerr. Parameters of the 
models are shown in Table~\ref{model}. In this Table, $\varepsilon_0$, 
$\omega_{\rm H}$ and $r_{\rm H}$ are the maximum energy density of the toroid, 
the value of the potential $\omega$ on the event horizon and the horizon 
radius, respectively.  In order to compare our models
with those of Kerr black holes, the values of $J/M^2$ and the ratios of the
radius of the marginally stable orbit to the horizon radius are
tabulated in Table~\ref{Kerr}. Here $R_{\rm ms}$ is the radius of
the marginally stable orbit.

%
\begin{table*}
\begin{minipage}{100mm}
\caption{Model parameters of toroid -- black hole systems}
\label{model}
\begin{tabular}{@{}lcccccccc}
Model & $q$ & $J\varepsilon_0$ & $M\varepsilon_0^{1/2}$ & 
$\omega_{\rm H}/\varepsilon_0^{1/2}$ & $J/M^2$ & 
$r_{\rm in}/r_{\rm H}$ & $r_{\rm out}/r_{\rm H}$ \cr
A & $1.2\times10^{-2}$ & $7.42\times10^{-6}$ & $6.23\times10^{-3}$ & $1.0\times10^1$ & $0.191$ & $2.04$ &  $5.56$ \cr
B & $8.0\times10^{-2}$ & $3.86\times10^{-4}$ & $2.34\times10^{-2}$ & $1.0\times10^1$ & $0.706$ & $1.37$ &  $4.59$ \cr
C & $4.9\times10^{-1}$ & $1.60\times10^{-3}$ & $4.61\times10^{-2}$ &  $4.0$          & $0.751$ & $1.46$ &  $5.06$ \cr
\end{tabular}
\end{minipage}
\end{table*}
%
%
\begin{table*}
\begin{minipage}{70mm}
\caption{Quantities for Kerr black holes}
\label{Kerr}
\begin{tabular}{@{}lccc}
$J/M^2$ & $R_{\rm ms}/r_{\rm H}$ & Comment        \cr
0.0   & 3.00 & Schwarzschild                      \cr
0.191 & 2.71 & $J/M^2$ is the same as Model A     \cr
0.50  & 2.27 &                                    \cr
0.706 & 1.97 & $J/M^2$ is the same as Model B     \cr
0.751 & 1.90 & $J/M^2$ is the same as Model C     \cr
0.998 & 1.16 &                                    \cr
1.00  & 1.00 & Extreme  Kerr                      \cr
\end{tabular}
\end{minipage}
\end{table*}

\subsection{Images of toroids}

\begin{figure*}
\begin{tabular}{cccc}
(a) $\theta_0=85\degr$&\epsfile{file=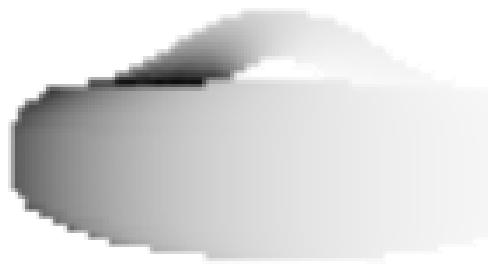,width=0.23\textwidth}&
(b) $\theta_0=60\degr$&\epsfile{file=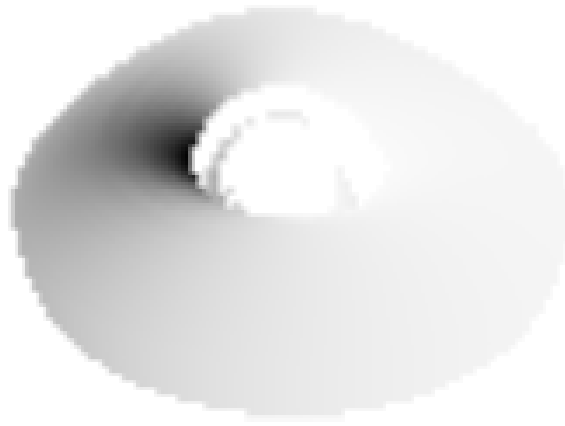,width=0.23\textwidth}\\
(c) $\theta_0=30\degr$&\epsfile{file=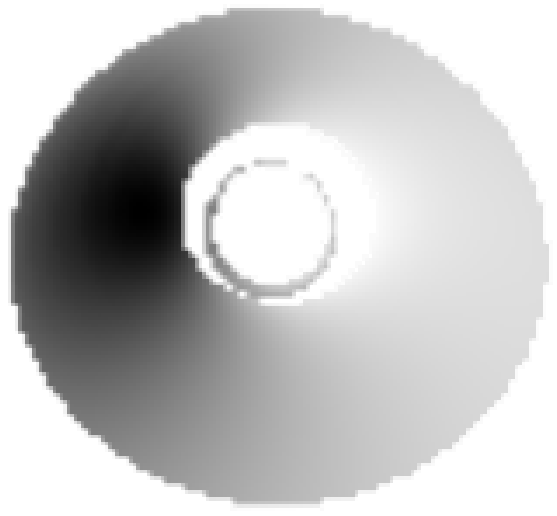,width=0.23\textwidth}&
(d) $\theta_0=5\degr$&\epsfile{file=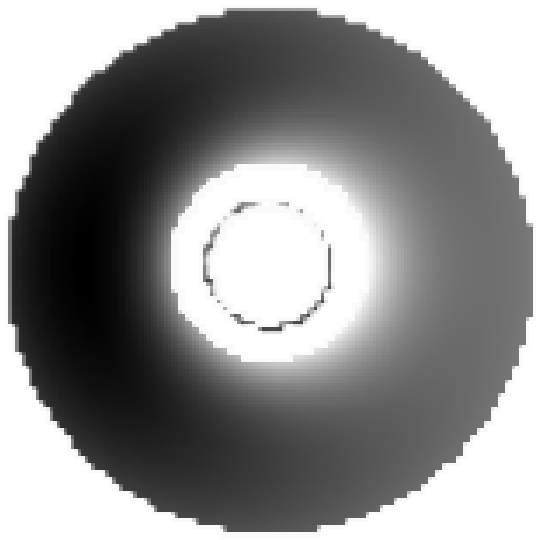,width=0.23\textwidth}\\
\end{tabular}
\caption{
Images of the accretion toroidal Model B for different inclination angles 
$\theta_0$. The shade of this image indicates the distribution of the total 
surface brightness or the total intensity. For Model B the mass ratio and 
the emissivity index are $q=0.08$ and $s=-1$, respectively. The inner radius 
$r_{\rm in}= 1.37 r_{\rm H}$ and the outer radius $r_{\rm out} = 4.59 
r_{\rm H}$. The toroid rotates counter-clockwise so that the left part of 
the toroid is approaching the observer.
}
\label{images}
\end{figure*}

Images of thin disks have been computed by several authors (e.g. Luminet 1978; 
Fukue \& Yokoyama 1988; Fanton et al. 1997). The characteristic feature of 
images of thin disks is significant deformation of disk shapes in the 
rear of black holes. Similar deformation can be seen even for 
self-gravitating thick toroids (see Figure \ref{images}).  

In this figure, images of self-gravitating toroids seen from different
inclination angles are shown.  The observer is assumed to be at 
$r_{\rm obs} = 10.5 r_{\rm H}$. This distance has been chosen because the 
numerically solved spacetime covers only the region with $r \le 2 r_{\rm out}$.
Although it seems too near to the black hole, photon trajectories beyond this 
distance are almost straight so that the images of the toroids and the line
profiles would be affected only little even if the observer is located
at a more distant position. 

\begin{figure*}
\epsfile{file=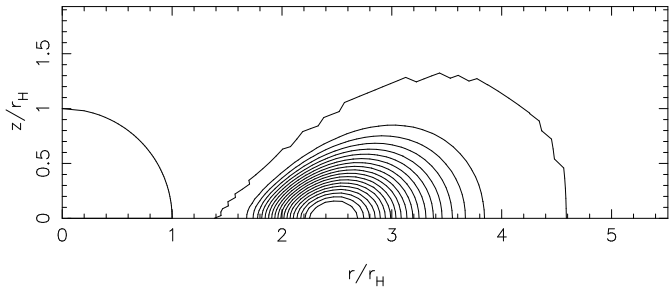,width=0.5\textwidth}
\caption{The contours of the energy density in a meridional cross section of
Model B. The circle at the origin 
denotes the horizon of the black hole in this coordinates. The units
of the distance is $r_{\rm H}$.}
\label{shapes}
\end{figure*}

Since the direction of the rotation of the toroid is counter-clockwise, the 
left part of the toroid is rotating toward the observer. In these pictures, 
the index of the emissivity is fixed to $s=-1$ and Model B is chosen, i.e. 
$ q = 0.08$. The shape and the equi-energy density contours of the toroidal 
Model B in the meridional cross section are shown in Fig. \ref{shapes}.

The toroid with $\theta_0=5\degr$ is observed almost from the direction of the 
rotation axis and shows almost a round disk shape. As the inclination angle is 
increased, the shape begins to be distorted by the gravitational lens effect. 
The front side of the toroid almost keeps its original shape, but the rear 
side is significantly warped.  As for the thickness of the toroid, the image 
of the rear side of the toroid is narrower compared to that of the front
side.  One more significant feature is the appearance of an inner ring image
inside the toroidal image.  This ring image is one of the multiple images
which are formed by photons orbiting several times around the black hole.
In particular, this ring is formed by 2nd or higher order images because
the 1st order image is hidden by the thickness of the toroid.

\subsection{Line profiles from self-gravitating toroids}

It is well known that the line profiles from disks or toroids
around black holes show characteristic features as discussed in Introduction:
1) broad line profiles, in particular towards the red wing and 2) asymmetry of 
the line profiles.  For thin disks, the profile varies as the inclination 
angle changes. When the inclination is small, $\theta_0 \la 15 \degr$, the 
line profile has a single peak around the wavelength or the frequency shifted 
from the intrinsic value by the amount of the gravitational redshift and the 
wing extends more to the red side compared to the blue wing. As the 
inclination becomes large, $\theta_0  \ga 15 \degr$, the peak shifts to the 
blue side because of the Doppler effect of the approaching region and 
the red wing extends further to the low frequency side also because of the 
Doppler shift of the receding inner region. At the same time, there appears a 
second peak in the red wing around the redshift factor which depends on the 
inclination angle and on the width of the disk. In other words, double-peaked 
feature, a strong blue peak and a weak red peak, appears. It is noted 
that positions of peaks are not affected by the types of the emissivity 
but depend on the inclination angle and on the width of thin disks. This 
is because the position of the peak is determined from the size of the area 
which has a same redshift factor on the surface of the toroid.  

\begin{figure*}
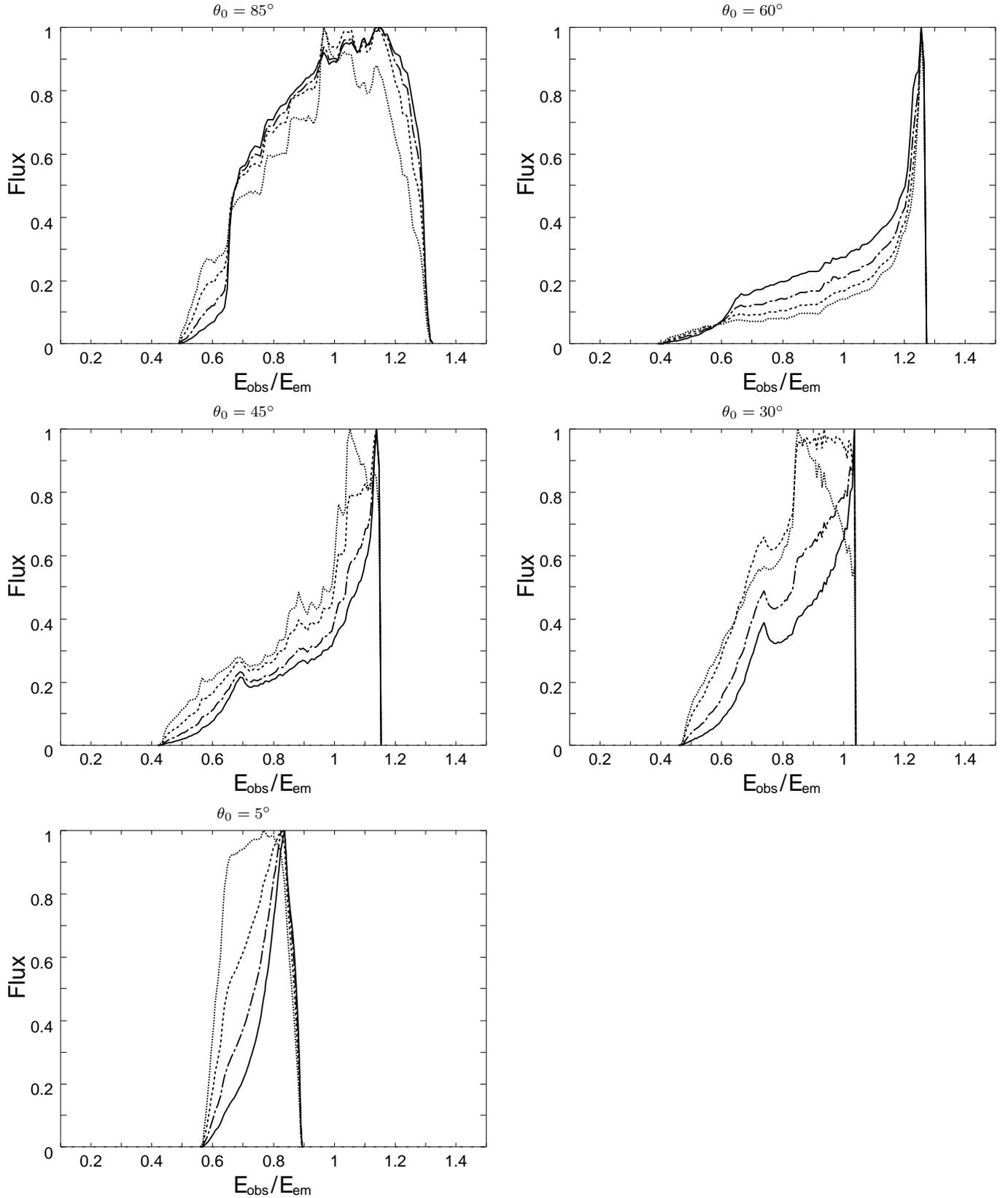

\begin{tabular}{cc}
$\theta_0=85^{\circ}$&$\theta_0=60^{\circ}$\\
\epsfile{file=power5.eps,width=0.5\textwidth}&
\epsfile{file=power30.eps,width=0.5\textwidth}\\
$\theta_0=45^{\circ}$&$\theta_0=30^{\circ}$\\
\epsfile{file=power45.eps,width=0.5\textwidth}&
\epsfile{file=power60.eps,width=0.5\textwidth}\\
$\theta_0=5^{\circ}$&\\
\epsfile{file=power85.eps,width=0.5\textwidth}&\\
\end{tabular}
\caption{
Line profiles of the accretion toroid (Model B, i.e. $q = 0.08$)
are shown for different inclination angles $\theta_0$ and for several 
emissivity indices $s$. Each flux is normalized so that its maximum 
becomes unity. The solid line, the dash-dotted line, the short dashed
line and dotted line denote models with the emissivity indices
$s = -1, -2, -3$ and $-4$, respectively.
}
\label{profile q-const}
\end{figure*}

\begin{figure*}
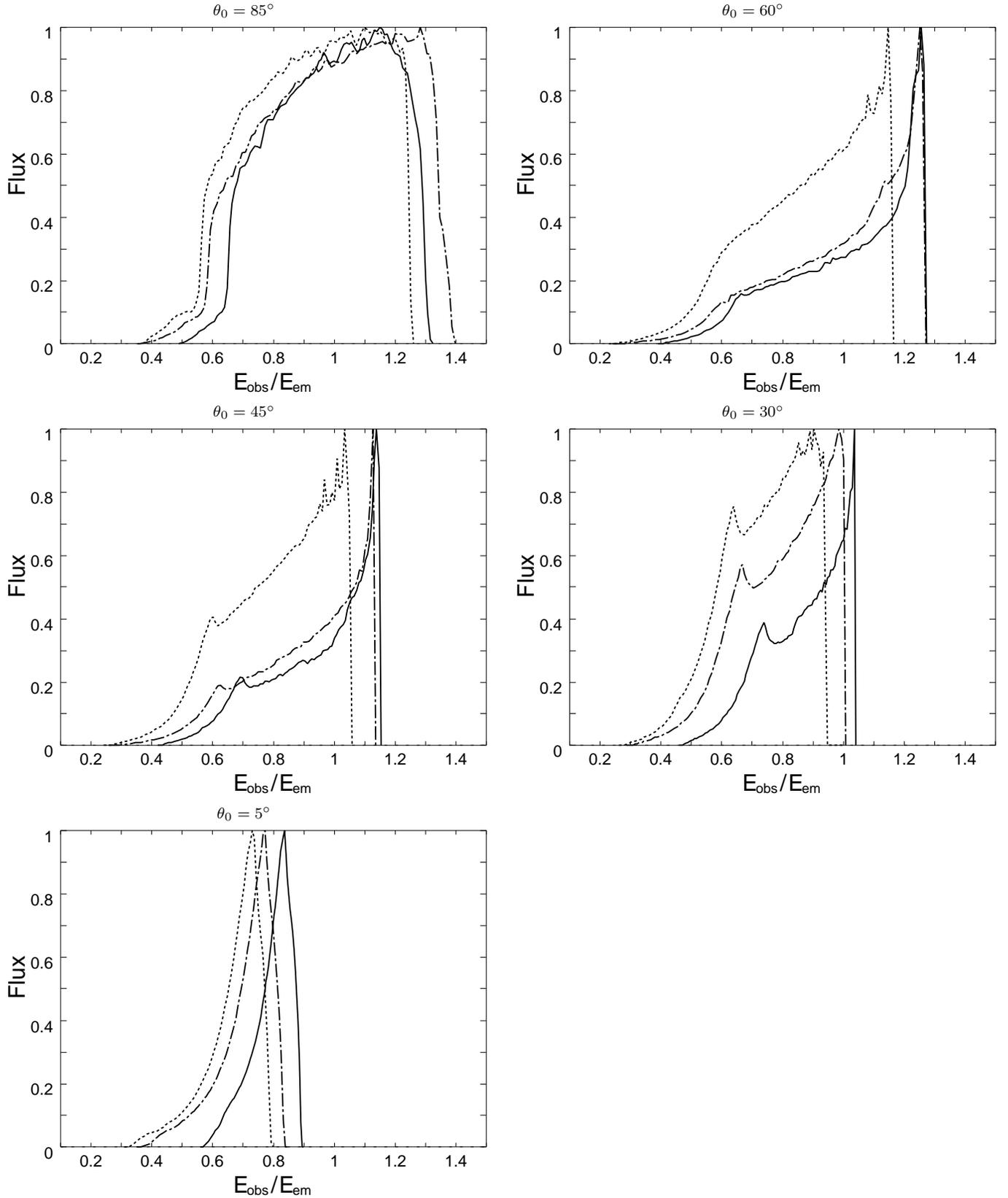

\begin{tabular}{cc}
$\theta_0=85^{\circ}$&$\theta_0=60^{\circ}$\\
\epsfile{file=mass5.eps,width=0.5\textwidth}&
\epsfile{file=mass30.eps,width=0.5\textwidth}\\
$\theta_0=45^{\circ}$&$\theta_0=30^{\circ}$\\
\epsfile{file=mass45.eps,width=0.5\textwidth}&
\epsfile{file=mass60.eps,width=0.5\textwidth}\\
$\theta_0=5^{\circ}$&\\
\epsfile{file=mass85.eps,width=0.5\textwidth}&\\
\end{tabular}
\caption{
Same as Fig. 4 but for Models A (solid line), B 
(dash-dotted line) and C (dotted line) with $s=-1$. 
}
\label{profile s-const}
\end{figure*}

Concerning the self-gravitating toroids, the line profiles are shown in 
Figs.~\ref{profile q-const} and \ref{profile s-const}. 
In Fig.~\ref{profile q-const}, the line profiles for Model B
with $q = 0.08$ are displayed for four different emissivity indices $ s = -1$, 
$-2$, $-3$ and $-4$. As seen from these figures, although basic features of 
line profiles are similar to those of thin disks, double-peaked structures
are no more pronounced even if they exist. This is mainly because a part of
the surface of the toroid is eclipsed by the toroid itself (Kojima \& Fukue 
1992; Bao \& Stuchlik 1992).  However, this self-eclipse does not lead to 
flat-top profiles obtained by Kojima \& Fukue~(1992) because of the general 
relativistic effect.

It is remarkable that the value of $g$ extends to 0.4 as seen from the profile 
for $\theta_0 = 60\degr$. In order to get this value of the redshift for the 
Kepler disk around a Kerr black hole, the value of $J/M^2$ is 0.5 or so 
(see e.g. Dabrowski et al. 1997) and the inner edge of the disk must be located
at the marginally stable orbit, i.e. $r_{\rm in} = 2.27 r_{\rm H}$. 
On the contrary, for our toroidal model, those values are $J/M^2 = 0.706$ 
and $r_{\rm in} = 1.27 r_{\rm H}$. This implies that the location of
the inner edge can be much nearer to the central black hole for {\it
toroidal} configurations compared to thin disk models in order to
get models with the same value of the maximum redshift.  
This occurs because the rotation law is not Keplerian so that the inner edge
of the toroid can be located nearer than the marginally stable radius
for the Kepler disks.  The similarity between our Model B
and the Kerr black hole -- Kepler disk model with $J/M^2 = 0.5$
can be seen in profiles for other moderate inclination angles apart
from the blue side.

In Fig.~\ref{profile s-const}, line profiles are shown for Models A--C with
different inclination angles and the emissivity index $s=-1$. 
The effect of the toroidal gravity to the spacetime can 
be clearly seen by the shift of the line to the red side for profiles with 
smaller inclination angles because the redshift for these models is mainly due 
to gravity. For self-gravitating toroids, the energy is shifted
significantly to the red side and the flux increases for every energy range.
For Model C, we have obtained $g_{\rm min} = 0.2$ and $J/M^2 = 0.75$,
where $g_{\rm min}$ is the minimum value of the quantity $g$.
In order to get $g_{\rm min} = 0.2$ for the Kerr model, we have to choose
a nearly extreme Kerr black hole with $J/M^2 \sim 1.0$ (see e.g. Dabrowski 
et al. 1997). It should be noted that the inner edge of the toroid is
located very near the central black hole, i.e. $r_{\rm in} = 1.46 r_{\rm H}$ 
for our toroidal model and $r_{\rm in} = 1.16 r_{\rm H}$ for the extreme Kerr 
case (see Table~\ref{Kerr}).

\section{DISCUSSION AND CONCLUSION}

In this paper we have constructed toroid -- black hole systems in general
relativity. We have computed photon trajectories in the numerically obtained 
gravitational fields and investigated the images of the toroids as well as the 
line profiles from the toroids. In these computations, we have assumed that 
the observer is located very near the black hole because the metric in the 
whole space has not been calculated. Therefore, the quantitative values in 
this paper will be changed a little if the observer is located at infinity. 
However, the important thing is not to obtain exact quantitative values but to
know characteristic feature which appears only by introducing thickness and 
self-gravity numerically exactly. In this sense, our results serve as 
representative ones for self-gravitating toroids around black holes.

\begin{figure*}
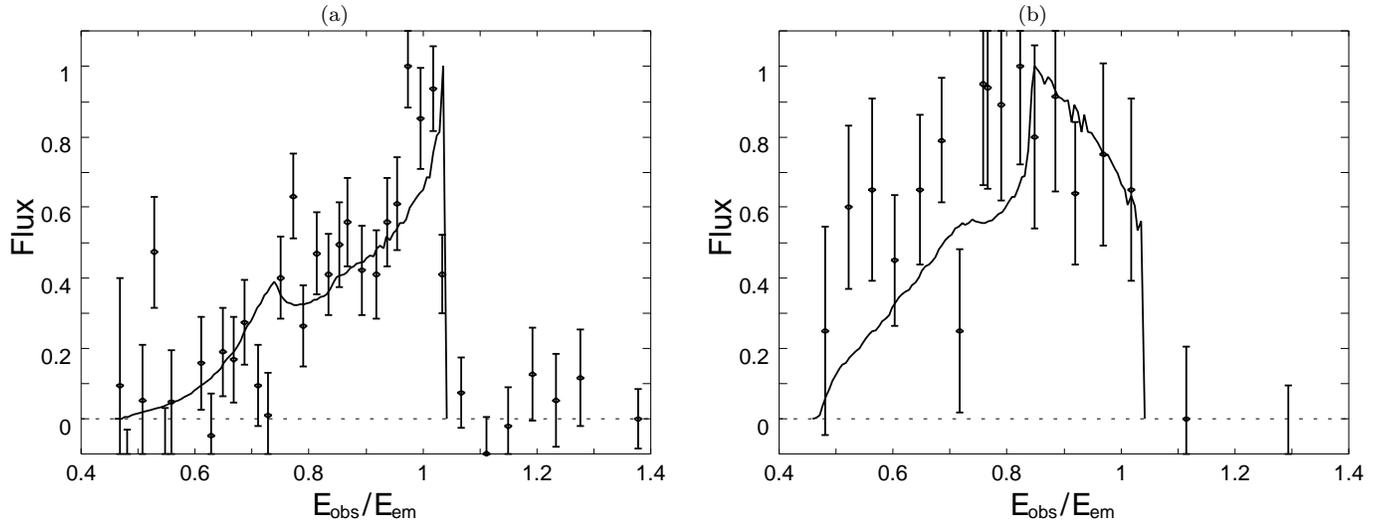

\begin{tabular}{cc}
(a)&(b)\\
\epsfile{file=tanaka.eps,width=0.5\textwidth}
&\epsfile{file=iwasawa.eps,width=0.5\textwidth}\\
\end{tabular}
\caption{
Same as Fig. 4 but for the model with $\theta_0 = 
30 \degr$ and emissivity indices (a) $s=-1$ (solid line) and (b) $s=-4$ 
(solid line). The observational data of the ASCA are plotted by points 
with error bars for the data of Tanaka et al.~(1995) (a) and that
of Iwasawa et al.~(1996) (b).
}

\label{comparison}
\end{figure*}

As discussed in Introduction, the observation of MCG--6--30--15 has given
us time varying line profiles whose interpretation has not been clarified yet
(Tanaka et al. 1995; Iwasawa et al. 1996; Dabrowski et al. 1997).
Some authors have proposed models for the system which would explain
the observations (Dabrowski et al. 1997; Bromley et al. 1997; Bromley et al. 
1998). However, since there are many parameters about the disk structures
and the X-ray sources, it is very difficult to obtain a unique solution
for the system.  This can be also seen from our result. For our Model B
with the emissivity indices $s = -1$ and $s = -4$, the line profiles are
shown in Fig.~\ref{comparison} together with the observational data. 
As seen from this figure, the tendency of the observational data seems to 
be roughly explained by the change of the emissivity index of
Model B.  Thus, at the present stage, we can only say that since 
the model cannot be uniquely determined, we have to get more accurate 
observational data to clarify the spacetime of the central region of
galaxies.

\section*{ACKNOWLEDGMENT}

We would like to thank Dr. A. Lanza for his discussion and his information 
about the paper in which self-gravitating thin disks are treated.

\end{document}